\input phyzzx.tex
\def\hilb{${\cal V}\otimes{\cal V}$}

\def\stackrel#1#2{\mathrel{\mathop{#2}\limits^{#1}}}

\def\dabliu{$W_{\infty}$}
\def\dablius{$W_{\infty}\times W_{\infty}$}
\Pubnum={$\rm UTS-DFT-93-11$}
\date={}
\pubtype={}
\titlepage
\title{QUANTUM DEFORMED CANONICAL TRANSFORMATIONS,\break~ 
$W_{\infty}$-ALGEBRAS~ AND UNITARY TRANSFORMATIONS}
\author{E.Gozzi$^{\flat}$ and M.Reuter$^{\sharp}$}
\address{$\flat$ Dipartimento di Fisica Teorica, Universit\`a di Trieste,\break
Strada Costiera 11, P.O.Box 586, Trieste, Italy \break and INFN, Sezione 
di Trieste.\break
\break
$\sharp$ Deutsches Elektronen-Synchrotron DESY,\break Notkestrasse 85, 
W-2000 Hamburg 52, Germany}
\abstract
We investigate the algebraic properties of the quantum counterpart of
the classical canonical transformations using the symbol-calculus
approach to quantum mechanics. In this framework we construct a set of
pseudo-differential operators which act on the symbols of operators,~\ie,
~on functions defined over phase-space. They act as operatorial left-
and right- multiplication and form a  ~$W_{\infty}\times W_{\infty}$-
algebra which contracts to its diagonal subalgebra in the classical
limit. We also describe the Gel'fand-Naimark-Segal (GNS) construction 
in this language and show that the GNS representation-space 
(a doubled Hilbert space) is closely related to the algebra of functions 
over phase-space equipped with the star-product of the symbol-calculus.

\endpage
\chapter{INTRODUCTION}
The Dirac canonical quantization\Ref\PAM{P.A.M.Dirac, "{\it The Principles of Quantum Mechanics}",~Oxford Univ.
Press,\nextline Oxford,~1958} can be
considered as a map\Ref\abra{P.Chernoff, Hadronic Jour.
4 (1981) 879;\nextline
R.Abraham and J.Marsden,~ "{\it
Foundations of Mechanics}", Benjamin, New York (1978)}\Ref\dunn{G.V.
Dunne, Jour.Phys.A 21 (1988) 2321}.
~"${\cal D}$"~sending real functions defined on phase space
~$f_{1},f_{2},\cdots$~to hermitian operators~${\widehat f}_{1},{\widehat f}_{2}
\cdots$~which act on some appropriate Hilbert space\refmark{2,3}
According to Dirac, this map should satisfy the following requirements
$$\eqalign{\ (a) & \ ~~ {\cal D}\bigl(\lambda_{1}f_{1}+\lambda_{2}f_{2}\bigr)=
\lambda_{1}{\widehat f}_{1}+\lambda_{2}{\widehat f}_{2},~~~\lambda_{1,2}\in R\cr
\ (b) & \ ~~ {\cal D}\bigl(\bigl\{f_{1},f_{2}\bigr\}_{pb}\bigr)={1\over i\hbar}
\bigl[{\widehat f}_{1},{\widehat f}_{2}\bigr]\cr
\ (c) & \ ~~ {\cal D}\bigl( 1\bigr)=I\cr
\ (d) & \ ~~ {\widehat q},{\widehat p}~~are~represented~irreducibly.
\cr}\eqn\unouno$$
Here ~$\bigl\{\cdot,\cdot\bigr\}_{pb}$~denotes the Poisson bracket, and ~$I$~is the 
unit operator.
\par
Groenwald and van Hove\Ref\VONHO{H.J.Groenwold, Physica 12 (1946) 405;\nextline
L. van Hove, Mem.de l'Acad.Roy. de Belgique  XXVI (1951) 61}. showed that it 
is impossible to satisfy
these conditions for all functions. They can be established only for 
functions which are at most quadratic in ~$p$~and~$q$. This obstruction to a 
complete quantization implies that not all canonical transformation
generated by an arbitrary real function ~$f$,~\ie,~$\delta (\cdot)=\bigl\{\cdot,
f\bigr\}_{pb}$~can be implemented as unitary transformations
$$\delta(\cdot)={1\over i\hbar}[(\cdot),{\widehat f}]\eqn\unodue$$
on the Hilbert space.
In order to avoid the consequences of this "no-go"~theorem one has to relax 
some of the conditions~of~\unouno. In geometric quantization
\Ref\woo{N.Woodhouse, "{\it Geometric Quantization}", Clarendon Press,
Oxford, 1980;\nextline
J.Sniatycki, "{\it Geometric quantization and quantum mechanics}",~Springer, 
New York, 1980},for instance,
condition ~$(d)$~is abandoned,  and in Moyal quantization
\Ref\moy{J.E.Moyal, Proc.Combridge Phil.Soc. 45 (1949) 99} condition
~$(b)$~is modified. Moyal quantization involves replacing the Poisson
bracket in ~$(b)$~by a new,~$\hbar$-dependent bracket, the Moyal bracket.
It is a "deformation" \Ref\flat{F.Bayen et al., Ann.of Physics 111 (1978) 61;
~ibid. 111 (1978) 111}
of the Poisson bracket to which it reduces in the classical limit~$\hbar
\rightarrow 0$.
\par
The formal structure underlying the Moyal quantization is the "symbol
calculus"\Ref\bere{F.A.Berezin, Sov.Phys.Usp.23 (1980) 763}. Generally
speaking it deals with the representation of operators acting on some Hilbert
space ~${\cal V}$~in terms of functions defined on a suitable manifold
~${\cal M}$. The "symbol map",~${symb}$, is a linear one-to-one map
from the space of operators on ~${\cal V}$~to ~$Fun({\cal M})$~,
the space of complex functions on ~${\cal M}$. We write ~$f=symb({\widehat f})$~
for the function ~$f$~ representing the operator~${\widehat f}$. The inverse
of the symbol map, the "operator map",~$op$, associates a unique operator to
any complex~$f\in Fun({\cal M})$~:~$op(f)={\widehat f}$. On ~$Fun({\cal M})$~one 
introduces the so-called star product~$\ast$~which expresses the operator
product ~${\widehat f}_{1}{\widehat f}_{2}$~in terms of the symbols of
~${\widehat f}_{1}$~and~${\widehat f}_{2}$: 
$$symb({\widehat f}_{1}{\widehat f}_{2})=symb({\widehat f}_{1})\ast symb
({\widehat f}_{2})\eqn\unotre$$
Like the operator product the star product is non-commutative but associative.
\par
Let us now return to quantum mechanics and let us assume that  ~${\cal V}$~
is the state space of some quantum mechanical system with ~$N$~degrees of
freedom and that ~${\cal M}\equiv{\cal M}_{2N}$~is its phase-space.
For simplicity we assume\ that ~${\cal M}_{2N}$~ is the flat
~$\Re^{2N}$.
A particularly important type of symbol is the Weyl symbol
\Ref\wewi{H.Weyl, Z.Phys.46 (1927) 1;\nextline
E.Wigner, Phys.Rev.40 (1932) 740}. It is defined as
$$f(\phi^{a})=\int{d^{2N}\phi_{0}\over (2\pi\hbar)^{N}}~exp\bigl[{i\over
\hbar}\phi_{0}^{a}\omega_{ab}\phi^{b}\bigr]Tr\bigl[{\widehat T}(\phi_{0})
{\widehat f}\bigr]\eqn\unoquattro$$
Its inverse is:
$${\widehat f}=\int {d^{2N}\phi~d^{2N}\phi_{0}\over (2\pi\hbar)^{2N}}
~f(\phi)~exp\bigl[{i\over\hbar}\phi^{a}\omega_{ab}\phi_{0}^{b}\bigr]
{\widehat T}(\phi_{0})\eqn\unosei$$
Here ~$\phi^{a}=(p^{1},\cdots p^{N},q^{1},\cdots q^{N})$,~$a=1\cdots 2N$,
are canonical coordinates on~${\cal M}_{2N}$,~and ~$\omega_{ab}$~are the 
coefficients of the corresponding symplectic two-form.
The operators~${\widehat T}(\phi_{0})$, sort of translation operators
on phase-space \Ref\litt{R.G.Littlejohn, Phys.Rep.138 (1986) 193}, 
are defined in ref.[15]. For the case when the phase-space is not
flat see ref.[11]\REF\STU{G.G.Emch, Jour.Math.Phys. 23 (1982) 1791.}.  
For real functions~$f$~the operator ~${\widehat f}$~is hermitian.
If ~$f=symb({\widehat f})$~is complex, then the complex-conjugate
function~$f^{\ast}$~is the symbol of the hermitian-adjoint
operator:
$$op(f^{\ast})={\widehat f}^{\dag}~~,~~f^{\ast}=symb({\widehat f}^{\dag})
\eqn\unoseia$$
Arbitrary (mixed) states of the system are characterized by
~$\varrho=symb({\widehat\varrho})$~where~${\widehat\varrho}$~is the density 
operator. In particular, for pure states~${\widehat\varrho}=\ket{\psi}
\bra{\psi}$~one obtains the familiar Wigner\refmark{9} function
$$\varrho_{\psi}(p,q)=\int d^{N}x~exp\bigl[-{i\over\hbar}px\bigr]
\psi(q+{1\over 2}x)\psi^{\ast}(q-{1\over 2}x)\eqn\unosette$$
Using the symbols of density operators~${\widehat\varrho}$~and
of observables~${\widehat{\cal O}}$~it is possible to reformulate the
full machinery of quantum mechanics and in particular to write down
the expectation values of observables and so on. Details about this are
provided in ref.[15].
The star product for the Weyl symbol is given by an infinite-order
differential operator:
$$\eqalign{\bigl(f\ast g\bigr)(\phi) \ = & \ f(p,q)~exp\bigl[{i\hbar\over 2}\bigl(
{\stackrel{\leftarrow}{\partial}\over\partial q^{i}}{\stackrel{\rightarrow}
{\partial}\over\partial p_{i}}-{\stackrel{\leftarrow}{\partial}\over
\partial p_{i}}{\stackrel{\rightarrow}
{\partial}\over\partial q^{i}}\bigr)\bigr]~g(p,q) \cr
\ = & \ f(\phi)~exp\bigl[i{\hbar\over 2}
\stackrel{\leftarrow}{\partial_{a}}\omega^{ab}
\stackrel{\rightarrow}{\partial_{b}}\bigr]g(\phi) \cr
\ = & \ \sum_{n=0}^{\infty}{1\over
n!}({i\hbar\over 2})^{n}\omega^{a_{1}b_{1}}\cdots\omega^{a_{n}b_{n}}
(\partial_{a_{1}}\cdots\partial_{a_{n}}f)(\partial_{b_{1}}\cdots\partial
_{b_{n}}g)\cr}\eqn\unonove$$
Here ~$\omega^{ab}$~is the inverse of ~$\omega_{ab}$:
$\omega^{ab}\omega_{bc}=\delta^{a}_{c}$. In the classical limit the star-product becomes the ordinary pointwise 
product:~$(f\ast g)(\phi)=f(\phi)g(\phi)+O(\hbar)$. 
By expressing either ~$f$~or ~$g$~in terms of its Fourier transform,
it is easy to derive other representations of the star product\refmark{8}.
The Moyal bracket~
$\bigl\{\cdot,\cdot\bigr\}_{mb}$~is defined as the commutator
with respect to this star multiplication:
$$\eqalign{\bigl\{f,g\bigr\}_{mb} \ = & \ {1\over i\hbar}\bigl(f\ast g-g\ast
f\bigr)\cr
\ = & \ symb\bigl({1\over i\hbar}[{\widehat f},{\widehat g}]
\bigr)\cr}\eqn\unodieci$$
For ~$\hbar\rightarrow 0$~the Moyal bracket approaches the Poisson bracket:
~$\{f,g\}_{mb}=\bigl\{f,g\bigr\}_{pb}+O(\hbar^{2})$. In our notation the
Poisson bracket reads
$\bigl\{f,g \bigr\}_{pb}\equiv\partial_{a}f~\omega^{ab}~
\partial_{b}g$.
From the fact that the star-multiplication is associative it follows that the
Moyal bracket fulfills the Jacobi identity and that
$$\bigl\{f,g_{1}\ast g_{2}\bigr\}_{mb}=\bigl\{f,g_{1}\bigr\}_{mb}\ast g_{2}+
g_{1}\ast\bigl\{f,g_{2}\bigr\}_{mb}\eqn\unododici$$
This means that ~$\bigl\{f,{\cdot}\bigr\}_{mb}$~with ~$f\in Fun({\cal M})$~
fixed, is a derivation\refmark{2} on the algebra ~
$\bigl(Fun({\cal M}),\ast\bigr)$.  For other properties regarding the
"derivation" property of the Moyal structure, we refer the reader to ref.[7].
\par 
Let us now replace the
Poisson bracket on the LHS of ~$(b)$~by the Moyal bracket. We can then find
a quantization map~${\cal D}$~which satisfies all 
the requirements ~$(a)-(d)$~of ~\unouno~{\it for all}
functions ~$f_{1,2}\in Fun({\cal M}_{2N})$, namely the operator map
~$op= symb^{-1}$:
$$op\bigl(\bigl\{f_{1},f_{2}\bigr\}_{mb}\bigr)={1\over i\hbar}\bigl[{\widehat
f}_{1},{\widehat f}_{2}\bigr]\eqn\unotredici$$
The unitary transformation
~$\delta_{f}(\cdot)={1\over i\hbar}\bigl[{\widehat f},(\cdot)\bigr]$~
are now in a one-to-one correspondence with the following
transformations acting on symbols
~$\delta_{f}(\cdot)  =  \bigl\{f,\cdot\bigr\}_{mb}= \bigl\{f,
\cdot\bigr\}_{pb}$~+higher~derivatives.~These 
transformations could be called
"{\it quantum deformed canonical transformations}". They
act on c-number functions, as in classical mechanics, but two
consecutive transformations are composed by forming the
Moyal bracket~$\bigl\{f_{1},f_{2}\bigr\}_{mb}$~instead of the
Poisson bracket.The purpose of this paper is to study the covariance properties
of quantum mechanics by using the notion of {\it quantum deformed
canonical transformations}. Our emphasis is on the relation between
the phase-space formulation and the standard Hilbert space formulation.
In Section-2 we construct a set of pseudo-differential
operators\Ref\pseu{L.H\"ormander, Comm.~on Pure and Applied
Math.  32 (1979) 359} on ~$Fun({\cal M}_{2N})$~corresponding to operatorial
left and right multiplication. They form a kind of ~$W_{\infty}\times W_{\infty}$~algebra. We give a careful discussion of its classical limit,
in which it contracts to the diagonal subgroup~$W_{\infty}^{diag}$.
In Section 3 we consider the deformed canonical transformations from the
point of view of the Gel'fand-Naimark-Segal (GNS) construction
\Ref\thir{I.Gel'fand and M.A.Naimark, Mat.Sborn., N.S. 12 [54] (1943) 197;\nextline
I.E.Segal, Ann.Math. 48 (1947) 930;\nextline
R.Haag, N.W.Hugenholtz and W.Winnink, Comm.Math. Phys. 5 (1967) 215;\nextline
for a review see :W.Thirring, "{\it Quantum Mechanics of Atoms and Molecules}",
Springer, New York, 1979}~which is at the heart of Thermo Field Dynamics
(TFD)\Ref\ter{I.Ojima, Ann.of Phys. 137 (1981) 1;\nextline
H.Umezawa, et al., {\it "Thermo field dynamics and condensed states"}, 
North-Holland, Amsterdam, 1982}, for instance. In this construction the doubled Hilbert space
~${\cal V}\otimes{\cal V}$~plays a central role. We show explicitly that
~\hilb~may be identified with the space of symbols,~$Fun({\cal M}_{2N})$.
In Section 4 we realize the~\dablius~algebra in the recently
proposed\Ref\noi{E.Gozzi, M.Reuter, "{\it Quantum Deformed Geometry
On Phase-Space}", \nextline  Mod. Phys.Lett. A, vol.8, no.15
 (1993) 1433;\nextline
E.Gozzi, M.Reuter, "{\it A Proposal for a Differential Calculus in Quantum
Mechanics}",~ Int.Jour.Mod.Phys.A in press.} extended Moyal formalism which was used to
formulate a tensor calculus on quantum phase-space.
\chapter{THE~\dablius~GENERATORS}
For each symbol ~$f= symb({\widehat f})$~we define the operators
~$L_{f}$~and ~$R_{f}$~acting on the space of symbols as
$$\eqalign{L_{f}g\ = & \ f\ast g=symb({\widehat f}~{\widehat g})\cr 
R_{f}g\ = & \ g\ast f = symb({\widehat g}~{\widehat f})\cr}\eqn\dueuno$$
for all ~$g=symb({\widehat g})$. The operators ~$L_{f}$~and
~$R_{f}$~implement on ~$Fun({\cal M}_{2N})$~the left and right
multiplication with ~${\widehat f}$, respectively. By employing
the form~\unonove~of the star-product, we can represent
~$L_{f}$~and ~$R_{f}$~by the following pseudo-differential
operators
$$\eqalign{L_{f} \ = & \ :f(\phi^{a}+{i\hbar\over 2}
\omega^{ab}\partial_{b}):\cr
R_{f} \ = & \ :f(\phi^{a}-{i\hbar\over 2}\omega^{ab}\partial_{b}):\cr}\eqn\duedue$$
Here the normal ordering symbol~$:(\cdots):$~means that all derivatives ~
$\partial_{a}$~should be placed to the right of all ~$\phi$'s.
This makes sure that
$$:f(\phi^{a}\pm{i\hbar\over 2}\omega^{ab}\partial_{b}): ~g(\phi)=
f(\phi^{a}\pm{i\hbar\over 2}\omega^{ab}{\partial\over\partial
{\widetilde\phi}^{b}})g({\widetilde\phi})\vert_{{\widetilde\phi}=\phi}\eqn
\duetre$$
The normal ordering can be performed explicitly in terms of an integral
involving the Fourier transform
~${\widetilde f}(l)=\int{d^{2N}\phi\over (2\pi)^{2N}}~exp~(-il_{a}
\phi^{a})f(\phi)$.
Since ~$:exp(il_{a}\phi{a}\pm {\hbar\over 2}l_{a}\omega^{ab}\partial_{b}):
=exp(il_{a}\phi^{a})~exp(\pm {\hbar\over 2}l_{a}\omega^{ab}\partial_{b})$~
one obtains
$$\eqalign{L_{f} \ = & \ \int d^{2N}l~{\widetilde f}(l)~
exp(il_{a}\phi^{a})~exp\bigl[-{\hbar\over 2}l_{a}\omega^{ab}
\partial_{b}\bigr]\cr
R_{f} \ = & \ \int d^{2N}l~{\widetilde f}(l)~exp(il_{a}\phi^{a})~exp
\bigl[+{\hbar\over 2}l_{a}\omega^{ab}\partial_{b}\bigr]\cr}\eqn\duecinque$$
Using the definition~\dueuno~and the fact that the star-product 
is associative, it is easy to work out the commutator algebra of the ~$L$'s
~and the ~$R$'s. \nextline For instance,
$\bigl[L_{f_{1}},L_{f_{2}}\bigr]g =   i\hbar L_{\{f_{1},f_{2}\}_{mb}}g$
and similarly for the other brackets. The result is
$$\eqalign{\ \bigl[ & \ L_{f_{1}},L_{f_{2}}\bigr]=+i\hbar 
L_{\{f_{1},f_{2}\}_{mb}}\cr
\ \bigl[ & \ R_{f_{1}},R_{f_{2}}\bigr]=-i\hbar R_{\{f_{1},f_{2}\}_{mb}}\cr
\ \bigl[ & \ L_{f_{1}},R_{f_{2}}\bigr]=0\cr}\eqn\duesette$$
We observe that the pseudo-differential operators ~$L_{f}$~ and ~$-R_{f}$~
generate two commuting copies ~${\cal A}_{L}$~and ~${\cal A}_{R}$~of the same
algebra ~${\cal A}$~which is isomorphic to the Moyal bracket algebra
on ~$Fun({\cal M}_{2N})$. In order to analyze the classical limit of the algebra
~\duesette~it is useful to define the following linear combinations
$$\eqalign{V_{f}\ = & \ {1\over i\hbar}(L_{f}-R_{f})\cr
A_{f}\ = & \ {1\over i\hbar}(L_{f}+R_{f})\cr}\eqn\dueotto$$
It is easy to see that they satisfy the commutation relations
$$\eqalign{\bigl[V_{f_{1}},V_{f_{2}}\bigr] \ = & \ V_{\{f_{1},
f_{2}\}_{mb}}~~~~~~~~(a)\cr
\bigl[V_{f_{1}},A_{f_{2}}\bigr] \ = & \ A_{\{f_{1},
f_{2}\}_{mb}}~~~~~~~~(b)\cr
\bigl[A_{f_{1}},A_{f_{2}}\bigr] \ = & \ V_{\{f_{1},f_{2}\}_{mb}}~~~~~~~~~(c)\cr}
\eqn\duenove$$
By a lengthy calculation it can be checked explicitly that the operators
~\duecinque~indeed satisfy this algebra
\Ref\fairie{D.B.Fairlie and C.K.Zachos, Phys.Lett.B 224 (1989) 101;\nextline
I.Bakas, Phys.Lett 228B (1989)57;~Com.Math.Phys.134 (1990)487;\nextline
I.M.Gelfand and D.B.Fairlie, Comm.Math.Phys. 136 (1991) 487}. In our approach 
we can avoid the lengthy calculation because we have 
related the operators ~\duecinque~to left and right multiplication.
For later use we also write down the exponential form of the generators:
$$\eqalign{\ op & \ \bigl(e^{L_{f}}g\bigr)=exp[{\widehat f}]~{\widehat g}\cr
\ op & \ \bigl(e^{R_{f}}g\bigr)={\widehat g}~exp[{\widehat f}]\cr
\ op & \ \bigl(e^{V_{f}}g\bigr)=  exp[-{i\over\hbar}{\widehat f}]~{\widehat g}
~exp[{i\over\hbar}{\widehat f}]\cr
\ op & \ \bigl(e^{A_{f}}g\bigr)=exp[-{i\over\hbar}{\widehat f}]~{\widehat g}
~exp[-{i\over\hbar}{\widehat f}]\cr}\eqn\duedieci$$
For real ~$f$, the operators ~$V_{f}$~generate unitary transformations~${\widehat g}
\rightarrow U{\widehat g}U^{-1}$, while the ~$A_{f}$'s map
~${\widehat g}\rightarrow U{\widehat g}U$. If, say, ~${\widehat g}=
{\widehat\varrho}=\ket{\psi}\bra{\psi}$~is the density matrix associated
to the pure state ~$\ket{\psi}$~, then clearly ~$V_{f}$~maps pure
states on pure ones, while ~$A_{f}$~does not.
\par
Going now back to the operators~$L_{f}$~and ~$R_{f}$, the interesting
point about them is that they provide a link between the Hilbert space
formulation of QM and the geometry of phase-space. As is well known,
CM was  formulated in the language of phase-spaces and their
classical differential geometry  while QM was framed in the language 
of Hilbert spaces and operators. Thanks to the symbol calculus, we are now
in a position to study both CM and QM within the same setting, namely
as operations on ~$Fun({\cal M}_{2N})$. The quantum deformed canonical
transformations have a dual interpretation therefore: in the
Hilbert space formalism they are unitary transformations, in the phase- 
space formalism they are a kind "distorted" volume preserving diffeomorphisms
on~${\cal M}_{2N}$. It is interesting to investigate the novel geometric
structures which the Hilbert space formalism of QM induces on
phase-space. Let us first look at the situation in CM. There any real function
~$f\in Fun({\cal M}_{2N})$~gives rise to a hamiltonian vector field
~$h_{f}^{a}=\omega^{ab}\partial_{b}f(\phi)$
which generates a canonical transformation. This is a transformation
,~$\phi^{\prime a}=\phi^{a}+\delta\phi^{a}~~,~~\delta\phi^{a}=-h^{a}(\phi)
$~ which preserves the symplectic two form:
~${\partial\phi^{\prime a}\over\partial\phi^{c}}{\partial\phi^{\prime b}
\over\partial \phi^{d}}\omega_{ab}=\omega_{cd}$.
Scalar functions ~$g$~ transform as
$$\delta_{f}g=\bigl\{f,g\bigr\}_{pb}=-h^{a}_{f}\partial_{a}g\eqn\duequattordici$$
In particular, the time evolution of classical probability densities
~$\varrho_{cl}(\phi,t)$~is a special canonical transformation generated by
~$h^{a}\equiv h^{a}_{H}=\omega^{ab}\partial_{b}H$, where ~$H$~is the 
Hamiltonian. Therefore
$$\partial_{t}\varrho_{cl}=\bigl\{H,\varrho\bigr\}_{pb}=-h^{a}\partial_{a}
\varrho_{cl}\eqn\duequindici$$
In QM the density operator evolves according to von Neumann's equation
$$i\hbar~\partial_{t}{\widehat\varrho}=\bigl[{\widehat H},{\widehat\varrho}
\bigr]\eqn\duesedici$$
which, in the symbol language, becomes
$$\partial_{t}\varrho=\bigl\{H,\varrho\bigr\}_{mb}\eqn\duesedicia$$
Here ~$\varrho=symb({\widehat\varrho})$~is similar to the classical
density~$\varrho_{cl}$~in ~\duequindici~, but it is not positive definite: 
it is referred to as a pseudodensity. A special example of
a pseudodensity is the Wigner function~\unosette. Using~\unodieci,~\dueuno
~and~\dueotto, eq.~\duesedicia~may be rewritten as
$$\partial_{t}\varrho={1\over i\hbar}\bigl(L_{H}-R_{H}\bigr)\varrho
=V_{H}\varrho\eqn\duediciasette$$
More generally, any transformation of the form
$$\delta_{f}{\widehat g}={1\over i\hbar}\bigl[{\widehat f},{\widehat g}
\bigr]\eqn\duediciotto$$
can be written as
$$\delta_{f}g=V_{f}g\eqn\duediciannove$$
For ~$f$~real, ~$op(f)={\widehat f}$~is hermitian so that~\duediciotto~
is indeed an infinitesimal unitary transformation~${\widehat g}\mapsto
{\widehat U}{\widehat g}{\widehat U}^{-1}$. Comparing~\duequattordici~
to ~\duediciannove~we see that the pseudo-differential operator
~$-V_{f}$~is the quantum deformed version of the hamiltonian vector field
~$h_{f}^{a}\partial_{a}$. A natural question to ask is which one, if any, is the
role of ~$A_{f}$~in classical geometry.
\par
Let us look at the classical limit ~$\hbar\rightarrow 0$~of the above
pseudo-differential operators in more detail. From eq.~\duedue~
we obtain
$$\eqalign{L_{f} \ = & \ f(\phi)-{i\hbar\over 2}h^{a}\partial_{a}+O(\hbar^{2})\cr
R_{f} \ = & \ f(\phi)+{i\hbar\over 2}h^{a}\partial_{a}+O(\hbar^{2})\cr
V_{f} \ = & \ -h^{a}_{f}\partial_{a}+O({\hbar})\cr
A_{f} \ = & \ {2\over i\hbar}f(\phi)+O(\hbar)\cr}\eqn\duediciasettea$$
We see that ~$L_{f}$,$R_{f}$, and ~$V_{f}$~start with a term of zeroth order
in~$\hbar$: both ~$L_{f}$~and ~$R_{f}$~are the usual pointwise multiplication
with ~$f(\phi)$~and ~~$V_{f}$~is a first order operator to lowest order.
The situation is different for ~$A_{f}$~which starts with a term of order
~${1\over\hbar}$. Therefore matrix elements of the form
~$<g_{1}\vert A_{f}\vert g_{2}>=\int d^{2N}\phi~g^{\ast}_{1}(\phi)~A_{f}~
g_{2}(\phi)$
blow up in the classical limit, and ~$A_{f}$~is not a well defined
map~$Fun({\cal M}_{2N})\rightarrow Fun({\cal M}_{2N})$\break anymore. We conclude that
only the transformations generated by ~$V_{f}$, but not those of ~$A_{f}$,
possess a classical limit. As a consequence the algebra~${\cal A}_{L}\times
{\cal A}_{R}$~of eq.~\duesette~or, equivalently,
of eq.~\duenove~is "contracted" in the classical limit to its
diagonal  subalgebra ~${\cal A}_{diag}$~generated by ~$V_{f}$:
~${\cal A}_{L}\times {\cal A}_{R}\longrightarrow{\cal A}_{diag}$. The remaining commutation relations
$$\bigl[V_{f_{1}}^{cl},V_{f_{2}}^{cl}\bigr]=V^{cl}_{\{f_{1},f_{2}\}_{pb}}\eqn
\duediciannovea$$
have the Poisson brackets appearing on the RHS now. Eq.~\duediciannovea~is the same as the
Lie bracket of the hamiltonian vector fields:
~$\bigl[h_{f_{1}}^{a}\partial_{a},h_{f_{2}}^{b}\partial_{b}\bigr]=
-h^{a}_{\{f_{1},f_{2}\}_{pb}}\partial_{a}$.
Again we see that it is very natural to consider~$V_{f}$~as the
{\it quantum deformed hamiltonian vector field}. For ~$\hbar\rightarrow 0$~
it becomes a first order operator,~\ie,~a generic vector field with a Lie 
bracket algebra. For ~$\hbar\ne 0$~this algebra is deformed to the algebra
~$(a)$~of~\duenove~which involves the Moyal bracket as "structure constants".
The deformed algebra cannot be represented by differential operators of
finite order but only by pseudo-differential operators,~\ie,
operators containing terms with an arbitrary number of derivatives
\foot{This is an indication of the non-local nature of quantum mechanics.}
$$V_{f}=\omega^{ab}\partial_{a}f~\partial_{b}-{\hbar^{2}\over 24}
\omega^{a_{1}b_{1}}\omega^{a_{2}b_{2}}\omega^{a_{3}b_{3}}~~\partial_{a_{1}}
\partial_{a_{2}}\partial_{a_{3}}f~~\partial_{b_{1}}\partial_{b_{2}}
\partial_{b_{3}}+\cdots\eqn\dueventuno$$
\par
Let us now come back to the algebra ~${\cal A}$~generated by ~$L_{f}$
or~$-R_{f}$. For the special case of a two-dimensional phase-space
~${\cal M}_{2}$,~${\cal A}$~can be identified with a version of the
~\dabliu-algebra found\refmark{16}~as the
~$N\rightarrow\infty$~limit of the ~$W_{N}$-algebras which arose in the 
studies of conformal field theory models\Ref\russ{A.B.Zamolodchikov, Theor.
Math.Phys.65 (1985) 1205;\nextline
A.B.Zamolodchikov, V.A.Fateev, Nucl.Phys.B280 [FS18] (1987) 644;\nextline
V.A.Fateev, S.L.Lykyanov, Int.Jour.Mod.Phys.A3 (1988) 507}. Recently
~\dabliu-algebras also appeared in the theory of planar fermion systems 
in the context of the fractional quantum Hall effect and of
matrix models\Ref\sak{S.Iso, D.Karabali, B.Sakita, Phys.Lett.B296 (1992) 143;
\nextline 
S.R.Das, A.Dhar, G.Mandal, S.Wadia,~Mod.Phys.~Lett.A7 (1992) 71,~937;\nextline
A. Cappelli, et al. Nucl.Phys.B386 (1993) 465; Phys.Lett.B306 (1993) 100\nextline
J.Fr\"ohlich and U.Studer, ETH-TH preprint;\nextline
J.Avan, A.Jevicki, Phys.Lett. B266 (1991) 35}.
An explicit form of our algebra~${\cal A}$~which displays its
structure constants is obtained by choosing a basis on ~$Fun({\cal M}_{2})$
and working out the Moyal brackets among the basis elements. We briefly
discuss the case of a phase-space with the topology\foot{Strictly speaking,
our previous discussion applies to ~${\cal M}_{2}=\Re^{2}$~only,
but its extension to  periodic boundary conditions is
straightforward.} of 
a torus\refmark{16}\Ref\flet{P.Fletcher, Phys.Lett.B248 (1990)~323}:
~${\cal M}_{2}=S^{1}\times  S^{1}$. The basis elements can be chosen as
\refmark{16}~$f_{\vec m}(\phi)=-exp({\vec m}\cdot {\vec\phi})
\equiv-exp (im_{a}\phi^{a})$
where ~${\vec m}\equiv (m_{1},m_{2})\in {\bf Z}^{2}$~and ~$0\leq\phi^{a}
\leq2\pi$. The Moyal brackets of the ~$f_{\vec m}$'s are easily calculated,
and the algebra~\duesette~becomes (with ~$L_{\vec m}\equiv L_{f_{{\vec m}}}$
~etc. and ~${\vec m}\omega{\vec n}\equiv m_{a}\omega^{ab}n_{b}$)
$$\eqalign{\bigl[L_{\vec m},L_{\vec n}\bigr] \ = & \ +2i\sin\bigl({\hbar\over
2}{\vec m}\omega{\vec n}\bigr)~L_{{\vec m}+{\vec n}}\cr
\bigl[R_{\vec m},R_{\vec n}\bigr] \ = & \ -2i\sin\bigl({\hbar\over 2}
{\vec m}\omega{\vec n}\bigl)~R_{{\vec m}+{\vec n}}\cr
\bigl[L_{{\vec m}},R_{{\vec n}}\bigr]\ = & \ 0\cr}\eqn\dueventitre$$
and similarly for ~\duenove:
$$\eqalign{\bigl[V_{\vec m},V_{\vec n}\bigr] \ = & \ {2\over\hbar}\sin\bigl(
{\hbar\over 2}{\vec m}\omega{\vec n}\bigr)~V_{{\vec m}+{\vec n}}\cr
\bigl[V_{\vec m},A_{\vec n}\bigr] \ = & \ {2\over\hbar}
\sin\bigl({\hbar\over 2}{\vec m}\omega{\vec n}\bigl)~A_{{\vec m}+{\vec n}}\cr
\bigl[A_{{\vec m}},A_{{\vec n}}\bigr] \ = & \ {2\over\hbar}\sin
\bigl({\hbar\over 2}{\vec m}\omega{\vec n}\bigl)~V_{{\vec m}+
{\vec n}}\cr}\eqn\dueventiquattro$$  
In the classical limit~$\hbar\rightarrow 0$~the first of these equations 
becomes
$$\bigl[V_{{\vec m}},V_{\vec n}\bigr]=({\vec m}\omega{\vec n})~V_{{\vec m}
+{\vec n}}\eqn\dueventicinque$$
This is the algebra of ~$sdiff(T^{2})$~of (classical) area preserving 
diffeomorphisms on the torus, sometimes referred to as the 
~$w_{\infty}$-algebra.
The ~$VV$-relations of ~\dueventiquattro~are a deformation of it:
the \dabliu-algebra. Note that the rescaled generators ~$-{i\over\hbar}
L_{\vec m}$~and ${i\over\hbar}R_{\vec m}$~satisfy the same commutation
relations. Therefore, at least for ~$\hbar\ne 0$,~we may also identify ~$
{\cal A}_{L}$~and ~${\cal A}_{R}$~with ~\dabliu. In a slight abuse of language, we shall call ~${\cal A}$~a ~\dabliu-
algebra even for dimensions~$2N>2$.~(See also ref.[20]\REF\fauo{D.B.Fairlie
et al., Phys.Lett.218B (1989) 203}~for~$2N$-dimensional lattice algebras.)
\chapter{THE GNS CONSTRUCTION}
Roughly speaking, the GNS construction\refmark{13} is a procedure
to go from trace-type averages to bra/ket-type averages.
Our discussion of
the GNS construction is informal and it has no pretension of
mathematical rigorosity; convergence and domain questions
will not be addressed here. Instead we shall emphasize the relation 
between the GNS framework and the symbol calculus.
Our presentation is close to the one of Ojima\refmark{14} and we restrict
ourselves to systems with finitely many degrees of freedom. For the much more
general case of systems with infinitely many degrees
of freedom we refer to the work of Haag, Hugenholtz and
Winnink  [13].
\par
Let us consider an arbitrary operator~${\widehat B}\in\Omega({\cal V})$~where~
$\Omega({\cal V})$\break denotes the space of operators on the Hilbert
space ~${\cal V}$. By choosing an orthonormal basis~
$\bigl\{\ket{\alpha}\bigr\}$~on  ~${\cal V}$~we can represent ~${\widehat
B}$~as
$${\widehat B}=\sum_{\alpha,\beta}~B_{\alpha\beta}\ket{\alpha}\bra{\beta}
~~,~~B_{\alpha \beta}\equiv <\alpha\vert{\widehat B}\vert\beta>\eqn\treuno$$
The GNS construction (at least in the version used in
Thermo Field Dynamics\refmark{14}) associates to this operator~${\widehat B}$~a vector
~$\Vert{\widehat B}\gg\in {\cal V}\otimes {\cal V}$~in the tensor
product of ~${\cal V}$~with itself:
$$\Vert{\widehat B}\gg=\sum_{\alpha,\beta}~B_{\alpha\beta}~\ket
{\alpha}\otimes\ket{\beta}\eqn\tredue$$
We denote the "vector" map relating ~$\Vert{\widehat B}\gg$~to~
${\widehat B}$~ as
$$vec~:~~~\Omega({\cal V})\rightarrow {\cal V}\otimes{\cal V}~~,~~
{\widehat B}\mapsto \Vert{\widehat B}\gg=vec({\widehat B})\eqn\treduea$$
This defines a one-to-one correspondence between the operators on ~${\cal V}$~
and the elements of the doubled Hilbert space
~${\cal V}\otimes{\cal V}$. On the other hand, the symbol map
$$symb~:~~~\Omega({\cal V})\rightarrow~Fun({\cal M}_{2N})~~,~~{\widehat B}
\mapsto symb({\widehat B})\eqn\treduebi$$
represents operators on ~${\cal V}$~by functions on ~${\cal M}_{2N}$.
As a consequence, the composite map ~$vec\circ symb^{-1}\equiv 
vec\circ op$~provides a one-to-one correspondence between the functions
on phase-space and the doubled Hilbert space:
~$Fun({\cal M}_{2N})\cong ~{\cal V}\otimes{\cal V}$. The rest of this section is devoted to a detailed study of this relation 
between functions on phase-space and states in the doubled Hilbert
space. As we shall see, this connection is rather intriguing from many
points of view.
\par
The original motivation\refmark{14}~for the introduction of
~${\cal V}\otimes{\cal V}$~was that it allows to represent {\it statistical}
averages  in a Hilbert space language.
In this way a state of a quantum system can be described by a vector in 
a Hilbert space,~${\cal V}\otimes{\cal V}$, even if this state,
from the point of view of ~${\cal V}$, is not a pure one.
This approach is particularly useful in quantum field theory at finite
temperature because, via the GNS construction, we can use all the
techniques we know from zero-temperature field theory (perturbation
theory, etc.) and apply them to the doubled theory "living" on
~${\cal V}\otimes{\cal V}$. This approach is known as the 
Thermo Field Dynamics\refmark{14}. In this framework states of the type
~$\ket{\psi}\otimes\ket{0}$~and ~$\ket{0}\otimes\ket{\psi}$~are
called "particles"~and "tilde-particles",~respectively. (Here ~$\ket{0}$~
is the vacuum state.) The usual attitude is that the tilde-particles are
only auxiliary quantitites without too much physical significance. 
In the light of the identification of ~$Fun({\cal M}_{2N})$~with
~${\cal V}\otimes{\cal V}$~we see that the "tilde
particles" are nothing unnatural, but rather something very important
because they allow for a close link between quantum mechanical 
structures and classical ones.
\par
Quite trivially we can associate to any operator ~${\widehat A}\in\Omega
({\cal V})$~a new operator acting on ~${\cal V}\otimes{\cal V}$,
namely ~${\widehat A}\otimes I$, where ~$I$~is the identity operator.
Let us calculate the expectation value of ~${\widehat A}\otimes I$~
in the state~$\Vert{\widehat B}\gg$~of ~\tredue. With the dual
vector defined as 
$$\ll{\widehat B}\Vert=\sum_{\alpha,\beta}~B^{\ast}_{\alpha\beta}
~\bra{\alpha}\otimes \bra{\beta}\eqn\trequattro$$
and using~$<\alpha\vert\beta>=\delta_{\alpha\beta}$,~one obtains
easily
$$\ll{\widehat B}\Vert~{\widehat A}\otimes I~\Vert{\widehat
B}\gg  =  Tr\bigl[{\widehat A}{\widehat B}{\widehat B}^{\dag}\bigr]
\eqn\trecinque$$
Let us assume that we are given a quantum system on ~${\cal V}$~with
a hermitian density operator which does not necessarily correspond
to a pure state (${\widehat\varrho}^{2}\ne {\widehat \varrho}$).
The eigenvalues ~$p_{\alpha}$~of ~${\widehat\varrho}$~give the
probability to find the system in the associated eigenstates
$\ket{\alpha}$. Then the trace averages
$$\VEV{{\widehat A}}=Tr\bigl[{\widehat A}~{\widehat\varrho}\bigr]\eqn\tresei$$
can be re-written as a bra-ket expectation value of the type ~\trecinque~once
we have found an operator~${\widehat B}\equiv{\widehat\varrho}^{1\over2}$~such that
~${\widehat B}{\widehat B}^{\dag}={\widehat\varrho}$. This operator is
easily constructed in the diagonal basis of ~${\widehat\varrho}$~in which the
diagonal elements are ~$B_{\alpha}=\bigl(p_{\alpha}\bigr)^{1\over 2}$.~Therefore
~${\widehat B}$~can be chosen hermitian. Thus
~$\VEV{{\widehat A}}=~\ll{\widehat\varrho}^{1\over 2}\Vert~
{\widehat A}\otimes I~\Vert{\widehat\varrho}^{1\over 2}\gg$
with
~$\Vert{\widehat\varrho}^{1\over 2}\gg=~\sum_{\alpha}~\bigl(p_{\alpha}\bigr)
^{1\over 2}~\ket{\alpha}\otimes\ket{\alpha}$~
in the eigenbasis of ~${\widehat\varrho}$. Even though we are dealing with 
mixed states, we have now succeeded in describing them by a vector in
a Hilbert space, rather than by a density operator, but the prize we have 
to pay is the doubling of the Hilbert space. 
\par
Let us now return to the Moyal formalism. So far we have the chain of
maps relating symbols to operators and states in ~${\cal V}\otimes{\cal V}$:
$$Fun({\cal M}_{2N}){{~op\atop \longrightarrow}\atop{\longleftarrow\atop symb}}
\Omega({\cal V}){{~vec\atop \longrightarrow}\atop{\longleftarrow\atop vec^{-1}}}
{\cal V}\otimes{\cal V}\eqn\treottob$$
It will be convenient to define the map
$$\Theta\equiv~vec\circ op~:~~~Fun({\cal M}_{2N})\rightarrow{\cal V}\otimes{\cal V}~,~~
B\rightarrow\Theta(B)=~\Vert{\widehat B}\gg\eqn\treottoa$$
which takes us directly from the symbols to the vectors in
~${\cal V}\otimes{\cal V}$. By their very definition, density operators have
a non-negative spectrum. They are represented by a subspace of ~
$Fun({\cal M}_{2N})$~which consists of symbols (pseudo-densities)
~$\varrho$~for which ~$op(\varrho)$~is non-negative. Unfortunately
it is not possible to characterize this subspace very explicitly
\Ref\balaz{A.Voros, Ann.~Inst. H. Poincar\'e 24 A (1976) 31; 26A (1977) 343}.
It follows from~\unosette~and the hermiticity
of ~${\widehat\varrho}$~that ~$\varrho$~is real. Similarly the
symbols of observables~${\cal O}={\cal O}^{\dag}\in\Omega({\cal V})$~
are real, but not subject to further restrictions. Nonhermitian
operators on ~${\cal V}$~have complex symbols in general. In order to 
investigate the meaning of complex conjugation on ~$Fun({\cal M}_{2N})$~
we introduce the following map~${\cal J}$ on any  of the three spaces
listed in ~\treottob. ~${\cal J}$~is defined as the {\it antilinear}
extension of the following operations:
$${\cal J}(f)=f^{\ast}~~~,~~~\forall f\in Fun({\cal M}_{2N})\eqn\trenove$$
$${\cal J}({\widehat B})={\widehat B}^{\dag}~~,~~~~~~~\forall {\widehat B}\in\Omega({\cal V})
\eqn\tredieci$$
$${\cal J}\bigl(\ket{\alpha}\otimes\ket{\beta}\bigr)=\ket{\beta}\otimes
\ket{\alpha}~~,~~\forall \ket{\alpha}\otimes\ket{\beta}
\in{\cal V}\otimes{\cal V}
\eqn\treundici$$
In the literature\refmark{13,14}~${\cal J}$~is referred to as the
modular conjugation operator. It plays an important role in the GNS
construction because, at the level of ~${\cal V}\otimes{\cal V}$,
it interchanges the first with the second Hilbert space, see eq.~\treundici.
For our purposes it is natural to introduce its action on symbols as ordinary
complex conjugation and on operators as hermitian conjugation
because then~${\cal J}$~commutes with ~$op$~ and ~$vec$:
$${\cal J}\bigl(vec(B)\bigr)={\cal J}\Vert{\widehat B}\gg=vec\bigl(
{\cal J}({\widehat B})\bigr)=vec\bigl({\widehat B}^{\dag}\bigr)=
\Vert{\widehat B}^{\dag}\gg\eqn\tredodici$$
$${\cal J}\bigl(op(B)\bigr)={\cal J}\bigl({\widehat B}\bigr)=op\bigl({\cal
J}(B)\bigr)=op\bigl(B^{\ast}\bigr)\eqn\tretredici$$
Eq.~\tredodici~follows easily from the definitions ~\tredue,~\treundici~
and in eq.~\tretredici~we used ~\unoseia.
Clearly ~${\cal J}$~is an involution, ~${\cal J}^{2}=1$. Because hermitian 
conjugation of operators changes the order of products, a similar 
property must hold for the symbols. In fact, from~\unonove~
it follows that
~${\cal J}\bigl(f\ast g\bigr)={\cal J}(g)\ast {\cal J}(f)$.
The modular conjugation ~${\cal J}$~is useful in pushing forward to
the space ~${\cal V}\otimes{\cal V}$~ the usual
operator  multiplication which takes place in the space ~${\cal V}$. 
An elementary calculation shows that for any ~${\widehat f}$,
${\widehat B}\in\Omega({\cal V})$
$$\Vert{\widehat f}{\widehat B}\gg=\bigl({\widehat f}\otimes I\bigr)
\Vert{\widehat B}\gg\eqn\trequindici$$
$$\Vert{\widehat B}{\widehat f}\gg={\cal J}\bigl({\widehat f}^{\dag}\otimes
I\bigr){\cal J}~\Vert{\widehat B}\gg\eqn\tresedici$$
The two ~${\cal J}$'s on the RHS of ~\tresedici~have the following origin:
in going from ~${\widehat B}$~in ~\treuno~to ~$\Vert{\widehat B}\gg$~in
~\tredue, we have turned the "bra"~$\bra{\alpha}$~into a "ket"
~$\ket{\alpha}$, therefore the {\it right}
multiplication of ~${\widehat B}$~by~${\widehat f}$~means that
~${\widehat f}$~acts on the second factor of
~${\cal V}\otimes{\cal V}$. In ~\tresedici~ this is achieved by first
interchanging the two factors of ~${\cal V}\otimes{\cal V}$, then acting with
~${\widehat f}^{\dag}$~on the first factor, and then interchanging the 
factors once more to restore the original order. In matrix language
we could write~${\cal J}\bigl({\widehat f}^{\dag}\otimes I\bigr){\cal J}
=I\otimes{\widehat f}^{T}$~
where ~$(\cdot)^{T}$~denotes transposition.
\par
We shall now return to the ~\dabliu-generators~$L_{f}$~and ~$
R_{f}$~ and use them in order to write
$$\eqalign{op\bigl(L_{f}B\bigr)\ = & \ {\widehat f}{\widehat B}\cr
op\bigl(R_{f}B\bigr) \ = & \ {\widehat B}{\widehat f}\cr}\eqn\trediciassette$$
Applying the ~$vec$-map to ~\trediciassette~and using ~\trequindici,~\tresedici
~yields for ~$\Theta\equiv vec\circ op$:
$$\eqalign{\Theta\bigl(L_{f}B\bigr) \ = & \ \bigl({\widehat f}\otimes I\bigr)
~\Theta\bigl(B\bigr)\cr
\Theta\bigl(R_{f}B\bigr)\ = & \ {\cal J}\bigl({\widehat f}^{\dag}\otimes I\bigl)
{\cal J}~\Theta\bigl(B\bigr)\cr}\eqn\trediciotto$$
These are the relations we wanted to derive. They show that, at the level 
of symbols, the operators~$L_{f}$~and ~$R_{f}$~induce transformations
on the first and second factor of the doubled Hilbert space
~${\cal V}\otimes{\cal V}$,~respectively. In practice ~$B$~could be the 
symbol of some positive operators~${\widehat B}\equiv 
{\widehat\varrho}^{1\over 2}$~representing a (mixed) state, say, 
and ~${\widehat f}\equiv A$~some hermitian observable
for instance. Then, in the language of Thermo Field
Dynamics\refmark{14}, the operator ~$A\equiv A\otimes I$~would refer to
the "particles"~while the 
"tilde operator"~${\widetilde A}\equiv {\cal J}A{\cal J}=
{\cal J}\bigl(A\otimes I\bigr){\cal J}$
would play the corresponding role for the "tilde particles" living in 
the second factor of ~${\cal V}\otimes{\cal V}$. Thus, for ~$f$~real,
~$L_{f}$~and ~$R_{f}$~generate unitary transformations on the Hilbert
spaces of "particles" and "tilde particles",~respectively. Eq.~\trediciotto~
yields
$$\eqalign{\Theta\bigl(e^{i L_{f}}B\bigr) \ = & \ \bigl(e^{i{\widehat f}}
\otimes I\bigr)\Theta(B)\cr
\Theta\bigl(e^{i R_{f}}B\bigr) \ = & \ {\cal J}\bigl(e^{-i{\widehat f}}
\otimes I\bigr){\cal J}~\Theta(B)\cr}\eqn\treventi$$
\par
In section 2 we saw that, at the quantum level, the canonical transformations
had a direct product structure~${\cal A}_{L}\times {\cal A}_{R}\sim
W_{\infty}\times W_{\infty}$. In the GNS language the transformations
correspond to independent unitary transformations on the first and the
second factor of the representation space, respectively. 
We have  argued that in the classical limit the algebra~
${\cal A}_{L}\times{\cal A}_{R}$~
is contracted to its diagonal subalgebra ~${\cal A}_{diag}$~generated
by~$V_{f}={1\over i\hbar}\bigl(L_{f}-R_{f}\bigr)$. These transformations
act on the first and the second factor of ~${\cal V}\otimes{\cal V}$~
with the same unitary transformation. In the case of the linear
combination~$A_{f}$,~which decouples for ~$\hbar\rightarrow 0$~,
"particles" and "tilde particles" would be transformed differently. 
Therefore only those quantum canonical transformations survive the
classical limit~$\hbar\rightarrow 0$~which treat particles and tilde 
particles on an equal footing. Thus ~${\cal A}_{f}$~ has no analogue in
the classical differential geometry of phase space.
\par
Looking now at the action of the modular conjugation~${\cal J}$~on~$L_{f}$,~$R_{f}$~(for ~$f$~real)
we see that they are are interchanged:
$$\eqalign{{\cal J}R_{f}{\cal J} \ = & \ L_{f}\cr
{\cal J}L_{f}{\cal J} \ = & \ R_{f}\cr
{\cal J}V_{f}{\cal J} \ = & \ V_{f}\cr
{\cal J}A_{f}{\cal J} \ = & \ -A_{f}\cr}\eqn\treventuno$$
This is most easily seen in eqs.\duedue. Eqs.\treventuno~
and the algebra ~\duenove~ suggest an amusing analogy with current
algebra in QCD,~say:~$R_{f}$~and ~$L_{f}$~are right- and left-handed currents
which are interchanged by the "parity operator"~${\cal J}$. The linear
combinations ~$V_{f}$~and ~$A_{f}$~correspond to vector and axial vector
currents which are even and odd under parity, respectively. The 
contraction to the vector-subgroup in the classical limit,
~${\cal A}_{L}\times {\cal A}_{R}\rightarrow{\cal A}_{diag}$, is reminiscent
of chiral symmetry breaking in this language. 
\par
Let us finally look at the time evolution of states in
~${\cal V}\otimes{\cal V}$~in more detail. In operatorial language
the time evolution of density operators~${\widehat\varrho}$~is
governed by von Neumann's equation
~$i\hbar~\partial_{t}{\widehat\varrho}=\bigl[{\widehat H},{\widehat
\varrho}\bigr]$~
which, in the symbol language, becomes
~$\partial_{t}\varrho=\bigl\{H,\varrho\bigr\}_{mb}$.
The ~$vec$-map would associate to ~${\widehat\varrho}$~the vector
~$\Vert{\widehat\varrho}\gg\in {\cal V}\otimes{\cal V}$, which is not
what we want however. Actually it is the bracket
~$\ll{\widehat\varrho}^{1\over 2}\Vert{\widehat A}\otimes I
\Vert{\widehat\varrho}^{1\over 2}\gg~=Tr\bigl[{\widehat\varrho}~{\widehat
 A}\bigr]$~
which gives the correct statistical average of ~${\widehat A}$.
This means that we have to distinguish the symbol of the density operator 
itself~$\varrho=symb\bigl({\widehat\varrho}\bigr)$
from the symbol representing its square root:
~$\sigma\equiv symb\bigl({\widehat\varrho}^{1\over 2}\bigr)$
Applying ~$symb$~to ~${\widehat\varrho}^{1\over 2}{\widehat\varrho}^{1\over 2}=
{\widehat\varrho}$~ we see that the symbol~$\sigma(\phi)$~is the "star-square
root" of ~$\varrho(\phi)$,\ie,
~$\sigma(\phi)\ast\sigma(\phi)=\varrho(\phi)$.
In the limit ~$\hbar\rightarrow 0$,~$\sigma_{cl}(\phi)=\varrho_{cl}
(\phi)^{1\over 2}$~is the ordinary square root of ~$\varrho_{cl}$.
Thus~$\Theta(\sigma)=vec\bigl(op(\sigma)\bigr)=\Vert{\widehat\varrho}
^{1\over 2}\gg$.
In order to reproduce the time-evolution of ~$\varrho$~we require that
$$\eqalign{\partial_{t}\sigma \  = & \ \bigl\{H,\sigma\bigr\}_{mb}\cr
\  = & \ {1\over i\hbar}\bigl(L_{H}-R_{H}\bigr)\sigma=V_{H}\sigma\cr}
\eqn\treventinove$$
because then eq.~\unododici~implies for ~$\varrho=\sigma\ast\sigma$~ the usual
time evolution~$\partial_{t}\varrho =  \bigl\{H,\varrho\bigr\}_{mb}$.
Applying ~$\Theta=vec\circ op$~to ~\treventinove~we obtain
the correponding equation\foot{A much more general derivation of this equation,
valid also for systems with infinte degrees of freedom, can be found
in H.Araki and E.J.Woods, Jour.Math.Phys. 4 (1963) 637.} for ~$\Vert{\widehat\varrho}^{1\over 2}\gg$:
$$\eqalign{i\hbar~\partial_{t}\Vert{\widehat\varrho}^{1\over 2}\gg \ = & \
\Theta\bigl(i\hbar~\partial_{t}\sigma\bigr)\cr
\ = & \ \Theta\bigl(L_{H}\sigma\bigr)-\Theta\bigl(R_{H}\sigma\bigr)\cr
\ = & \ \biggl[\bigl({\widehat H}\otimes I\bigr)-{\cal J}\bigl({\widehat H}
\otimes I\bigr){\cal J}\biggr]\Vert{\widehat\varrho}^{1\over 2}\gg\cr
\ = & \ {\bar H}\Vert{\widehat\varrho}^{1\over 2}\gg\cr}\eqn\tretrentuno$$
In the last line of eq.~\tretrentuno~we used~\trediciotto.
In eq.~\tretrentuno~we recognize\refmark{14} the Hamiltonian for the time evolution of
GNS states:
$${\bar H}\equiv {\widehat H}\otimes I-{\cal J}\bigl({\widehat H}\otimes I\bigr
){\cal J}\eqn\tretrentadue$$
Of course we could have derived ~\tretrentuno~in the standard manner by
starting directly from the equation for the operator~${\widehat\varrho}$.
Here instead we wanted to emphasize that it is the symbol
~$\sigma$~(rather than~$\varrho$)~which represents the GNS states in
~$Fun({\cal M}_{2N})$.~At the level of ~$\sigma$, the two parts of the Hamiltonian
~${\bar H}$ are the~$W_{\infty}$-generators ~$L_{H}$~
and~$R_{H}$:
$$\eqalign{H\equiv {\widehat H}\otimes I\ \longleftrightarrow & \ L_{H}\cr
{\widetilde H}\equiv {\cal J}\bigl({\widehat H}\otimes I\bigr){\cal J}
\ \longleftrightarrow & \ R_{H}\cr
{\bar H}\equiv H-{\widetilde H} \ \longleftrightarrow & \ i\hbar V_{H}\cr}
\eqn\tretrentatre$$
At this point one could ask why, even at the quantum level, the evolution 
of the symbols~$\sigma$~or ~$\varrho$,~respectively, involves 
only the operator ~$V_{H}$, but not ~$A_{H}$:
$$\partial_{t}\varrho(\phi,t)=V_{H}\varrho(\phi,t)\eqn\tretrentaquattro$$
Clearly the ~$V_{H}$-transformation corresponds to the usual
unitary time-evolution\break ${\widehat\varrho}(t)=U(t){\widehat\varrho}(0)
U^{-1}(t)$, but the ~$A_{H}$~transformation (remembering eq.~\duedieci)
would lead to~${\widehat\varrho}(t)=U(t){\widehat\varrho}U(t)$. Now let us
assume that, at ~$t=0$,~${\widehat\varrho}$~describes a pure state,~\ie,~
that ~${\widehat\varrho}(0)^{2}={\widehat\varrho}(0)$. This condition is 
preserved under the transformation generated by ~$V_{H}$, 
but it is not preserved by
~$A_{H}$. This means that ~$V_{H}$~ maps pure states on pure states, but
~$A_{H}$~maps pure states into mixed ones. It is important to see that
in the present formalism the evoluton of pure states to pure states is
protected by a {\it symmetry}, namely modular conjugation
~${\cal J}$. In fact, let us assume that we modify eq.~\tretrentaquattro~
by adding a piece containing an ~$A$-type generator:
$$\partial_{t}\varrho(\phi,t)=\bigl[V_{H}+\epsilon A_{G}\bigr]~\varrho
(\phi,t)\eqn\tretrentacinque$$
Here ~$\epsilon$~is a (small) real parameter and ~$G$~ a second Hamiltonian.
Applying ~${\cal J}$~ to eq.~\tretrentacinque, the operator inside the square
brackets changes to ~$V_{H}-\epsilon A_{G}$~according to~\treventuno,
so that eq.~\tretrentacinque~violates modular conjugation symmetry.
\par
We see that QM has a universal symmetry, modular conjugation,
which forbids pure states to evolve into mixed ones. It cannot be represented by
the standard formulation of QM as an automorphism of ~${\cal V}$, but this is 
possible in the space~${\cal V}\otimes{\cal V}$~which translates into
~$Fun({\cal M}_{2N})$~in
the Moyal approach formulated here. One of the reasons of why this formulation
is very attractive is that, as ~${\cal J}$~is now represented on the space
of states, it is on the same logical footing as the other (discrete) symmetries
we are familiar with. One could now try to find mechanisms that break
this symmetry and allow for the transition of pure states into mixed ones.

\chapter{THE EXTENDED MOYAL FORMALISM}
In this section we identify the ~${\cal A}_{L}\times{\cal A}_{R}$~
algebra in the extended Moyal formalism proposed recently\refmark{15}.
It was introduced in order to formulate a quantum deformed exterior
calculus on phase-space. The basic idea is to work on a~$8N$-dimensional
supermanifold~${\cal M}_{8N}$, the extended phase-space, which is closely 
related to the tangent and cotangent bundle over the standard phase-space
~${\cal M}_{2N}$. The coordinates on ~${\cal M}_{8N}$~ are the ~$8N$-tuples
~$(\phi^{a},\lambda_{a},c^{a},{\bar c}_{a})$~where ~$\phi^{a}$~are the usual
coordinates on ~${\cal M}_{2N}$,~$\lambda_{a}$~are commuting auxiliary 
variables, and ~$c^{a}$,~${\bar c}_{a}$~are  anticommuting coordinates. Standard
phase-space is identified with the ~$(\phi^{a},0,0,0)$-hypersurface
in ~${\cal M}_{8N}$. For functions on the extended phase-space, ~$A,B\in Fun({\cal M}_{8N})$~
one can introduce the extended star product~$\ast_{e}$:
$$A\ast_{e}B=A~exp\bigl[{i\over 2}\bigl({\stackrel{\leftarrow}
{\partial}\over\partial\phi^{a}}{\stackrel{\rightarrow}{\partial}
\over\partial\lambda_{a}}-{\stackrel{\leftarrow}{\partial}\over
\partial\lambda_{a}}{\stackrel{\rightarrow}{\partial}\over\partial\phi^{a}}
\bigr)+{\stackrel{\leftarrow}{\partial}\over\partial c^{a}}{\stackrel
{\rightarrow}{\partial}\over\partial {\bar c}_{a}}\bigr]B\eqn\quattrouno$$
and the extended Moyal  bracket~$(emb)$:
~$\bigl\{A,B\bigr\}_{emb}={1\over i}\bigl[A\ast_{e}B- (-)^{[A][B]}B\ast_{e}
A\bigr]$.Here ~$[A]=0$~or~$1$~depending on the Grassmann parity of ~$A$. Apart from
obvious grading factors, the extended Moyal bracket enjoys the same
algebraic properties as the ordinary Moyal bracket. 
In the extended Moyal brackets~$\lambda_{a}$~plays the role of
a "momentum"\refmark{15} conjugate to~$\phi^{a}$ and differently from the standard
Moyal bracket for which~$\bigl\{\phi^{a},\phi^{b}\bigr\}_{em}=\omega^{ab}$,
the ~$\phi^{a}$'s have vanishing ~$em$-brackets among themselves. They behave
like ~$2N$-position coordinates  on a ~$4N$-dimensional (bosonic)
phase-space ~${\cal M}_{4N}\equiv\{(\lambda_{a},\phi^{a})\}$. In fact, the
bosonic piece of~\quattrouno~follows from ~\unonove~by replacing 
~$q^{i}\rightarrow\phi^{a}$,~$p_{i}\rightarrow \lambda_{a}$.
Under diffeomorphisms on ~${\cal M}_{2N}$,~$\lambda_{a}$~and ~${\bar c}_{a}$
~transform like derivatives ~$\partial_{a}$~and ~$c^{a}$~like the 
differentials ~$d\phi^{a}$.
This allows us to represent differential forms on ~${\cal M}_{2N}$~
in terms of scalar functions on ~${\cal M}_{8N}$. A p-form ~$F=
{1\over p!}F_{a_{1}\cdots a_{p}}(\phi)
~d\phi^{a_{1}}\wedge\cdots\wedge d\phi^{a_{p}}$, say, is replaced by
$${\widehat F}={1\over p!}~F_{a_{1}\cdots a_{p}}(\phi)~c^{a_{1}}\cdots
c^{a_{p}}~\in Fun({\cal M}_{8N})\eqn\quattroquattro$$
The anticommutativity of the ~$c^{a}$'s mimics the wedge product
now. In the quantum deformed tensor calculus proposed in ref.[15],
operations such as the exterior derivative, the contraction or
the Lie derivative of ~$F$~were expressed as the extended
Moyal bracket of ~${\widehat F}$~with appropriate functions on
~${\cal M}_{8N}$. The Lie derivative\foot{For the definition
of the classical analog of this and similar operations, we refer the reader 
to refs.[2] and [22]\REF\uff{E.Gozzi, M.Reuter,
W.D.Thacker, Phys.Rev.D40 (1989) 3363}.} along the Hamiltonian vector
field ~$h$,~say, is given by
$$L_{h}{\widehat F}={\cal P}\bigl\{{\widehat F},{\widetilde{\cal H}}\bigr\}
_{emb}\eqn\quattroquattroa$$
Here~${\widetilde{\cal H}}$~is a certain "super-Hamiltonian"\refmark{22}
on the extended phase-space which is constructed from ~$h$. Furthermore
~${\cal P}$~is the projection operator
on the ~$(\lambda_{a}=0,{\bar c}_{a}=0)$-hypersurface in ~${\cal M}_{8N}$.
This type of exterior calculus has the property that "physical"
fields live in ~$(\phi,c)$-space, but the tensor manipulations are
realized by Moyal brackets on the larger space
~${\cal M}_{8N}$. This is why we need the projector ~${\cal P}$~in eq.
\quattroquattroa. We are not going into further details here; the interested
reader is referred to refs.[15]. Instead we shall give a self-contained 
description of the ~$W_{\infty}\times W_{\infty}$~generators on the ~
$\lambda,\phi$-space, 
~${\cal M}_{4N}$. We shall also identify the
modular transformations ~${\cal J}$~and the complex structure with
respect to which it is a conjugation. This is interesting in itself and we 
shall not consider the fermionic variables ~$c^{a}$~and ${\bar c}_{a}$
here.
\par
We consider the algebra of functions~$Fun({\cal M}_{4N})$~
equipped with the extended star-product~\quattrouno~with the
Grassmannian piece omitted. Loosely speaking,
we can identify the space ~${\cal M}_{4N}$~with the
tangent bundle\refmark{15},~$T{\cal M}_{2N}$, over standard phase-space.
The reason is that for any function~$f$~ on~${\cal M}_{2N}$~we have\break
~$\{f(\phi),\lambda_{a}\}_{emb}=\partial_{a}f(\phi)$. This means that 
in the Hamiltonian formalism, ~$\lambda_{a}$~plays the role of the
derivatives ~$\partial_{a}$,~\ie,~of a basis in tangent space. In a sense, what
we are considering here is the "lift" of the Moyal formalism from phase-space 
to its tangent bundle:~$\phi^{a}$~are coordinates in the base manifold
~${\cal M}_{2N}$~and ~$\lambda_{a}$~are coordinates in the fibers. In order
to find the relevant ~${\cal A}_{L}\times{\cal A}_{R}$~generators, it is 
advantageous to perform a change of variables which mixes the ~$\phi$'s
with the ~$\lambda$'s:
$$\eqalign{Z_{+}^{a} \  = & \ \phi^{a}+{\hbar\over 2}\omega^{ab}\lambda_{b}\cr
Z_{-}^{a} \ = & \ \phi^{a}-{\hbar\over 2}\omega^{ab}\lambda_{b}\cr}
\eqn\quattrocinque$$ 
When expressed in terms of these new variables the extended star-product 
becomes
$$A\ast_{e}B=A~exp\biggl[{i\hbar\over 2}\bigl({\stackrel{\leftarrow}{\partial}
\over\partial Z_{-}^{a}}\omega^{ab}{\stackrel{\rightarrow}{\partial}\over
\partial Z_{-}^{b}}-{\stackrel{\leftarrow}{\partial}
\over\partial Z_{+}^{a}}\omega^{ab}{\stackrel{\rightarrow}{\partial}\over
\partial Z_{+}^{b}}\bigr)\biggr]B\eqn\quattrosei$$
where ~$A$~and ~$B$~ are functions of both ~$Z_{+}^{a}$~and ~$Z_{-}^{a}$.
Let us look in particular at functions which depend on ~$Z_{-}^{a}$~or
~$Z_{+}^{a}$~only; we shall refer to them, with a slight abuse of
language, as "holomorphic" and "anti-holomorphic",~respectively.
Comparing eqs.~\quattrosei~and ~\unonove~we see that for (anti-) holomorphic
functions the extended star product has the same form as the usual one. In fact,
we can write
$$\eqalign{A(Z_{-})\ast_{e}~B(Z_{-}) \ = & \ A(Z_{-})~exp~\biggl[{i\hbar\over 2}
{\stackrel{\leftarrow}{\partial}\over\partial Z_{-}^{a}}\omega^{ab}{\stackrel
{\rightarrow}{\partial}\over\partial Z_{-}^{b}}\biggr]B(Z_{-})\cr
\ = & \ A(\phi)\ast B(\phi)\vert_{\phi=Z_{-}}\cr
A(Z_{+})\ast_{e} B(Z_{+})\ = & \ A(Z_{+})~exp~\biggl[-{i\hbar\over 2}
{\stackrel{\leftarrow}{\partial}\over\partial Z_{+}^{a}}\omega^{ab}{\stackrel
{\rightarrow}{\partial}\over\partial Z_{+}^{b}}\biggr]B(Z_{+})\cr
\ = & \ B(\phi)\ast A(\phi)\vert_{\phi=Z_{+}}\cr}\eqn\quattrosette$$
We also see that the ~$\ast_{e}$-product of a holomorphic with an 
anti-holomorphic function does not involve any derivatives:
$$\eqalign{A(Z_{-})\ast_{e}B(Z_{+}) \ = & \ A(Z_{-})B(Z_{+})\cr
A(Z_{+})\ast_{e}B(Z_{-}) \ = & \ A(Z_{+})B(Z_{-})\cr}\eqn\quattrootto$$
Similarly one finds for the extended Moyal brackets of (anti-)holomorphic
functions
$$\eqalign{\bigl\{A(Z_{-}),B(Z_{-})\bigr\}_{emb}\ = & \ +\hbar\bigl\{A(\phi),
B(\phi)\bigr\}_{mb}\vert_{\phi=Z_{-}}\cr
\bigl\{A(Z_{+}),B(Z_{+})\bigr\}_{emb}\ = & \ -\hbar\bigl\{A(\phi),
B(\phi)\bigr\}_{mb}\vert_{\phi=Z_{+}}\cr
\bigl\{A(Z_{-}),B(Z_{+})\bigr\}_{emb}\ = & \ 0\cr}\eqn\quattronove$$
We see that, with respect to the extended Moyal bracket, the holomorphic and
anti-holo-\break morphic functions form closed sub-algebras which are 
mutually commuting, and each of which is isomorphic to the standard 
Moyal bracket algebra on ~${\cal M}_{2N}$. In particular the ~$Z$'s
satisfy
$$\eqalign{\bigl\{Z_{-}^{a},Z_{-}^{b}\bigr\}_{emb} \ = & \ +\hbar\omega^{ab}\cr
\bigl\{Z_{+}^{a},Z_{+}^{b}\bigr\}_{emb} \ = & \ -\hbar\omega^{ab}\cr
\bigl\{Z_{+}^{a},Z^{b}_{-}\bigr\}_{emb} \ = & \ 0\cr}\eqn\quattrodieci$$
Apart from the factors of~$\pm\hbar$, we get two closed algebras similar
to~$\bigl\{\phi^{a},\phi^{b}\bigr\}_{mb}=\omega^{ab}$. This means that ~
$Fun({\cal M}_{4N})$,~equipped with the ~$\bigl\{{\cdot},{\cdot}\bigr\}_{emb}$
~is equivalent to two copies of ~$Fun({\cal M}_{2N})$~with 
the ordinary Moyal bracket. As we shall see, the modular conjugation
~${\cal J}$~interchanges the two copies.
\par
In order to recover the ~$W_{\infty}\times W_{\infty}$~generators on
~${\cal M}_{4N}$,~let us fix a real function \break $f\in Fun({\cal M}_{4N})$~
and let us define the following operators on ~$Fun({\cal M}_{4N})$:
$$\eqalign{{\cal L}_{f} \ = & \ f(Z_{-})\ast_{e}\cr
{\cal R}_{f} \ = & \ f(Z_{+})\ast_{e}\cr}\eqn\quattroundici$$
The notation ~$f=f(Z_{-}^{a})$~means  that we replace
~$\phi^{a}$~by~$Z_{-}^{a}$~in the original
~$f=f(\phi)$. In this way the operators~${\cal L}_{f}$~and
~${\cal R}_{f}$~have a non-trivial action on any (not necessarily
holomorphic and anti-holomorphic) function of ~$\phi$~and ~$\lambda$.
Using the associativity of the ~$\ast_{e}$-product and eqs.~\quattronove,
it is easy to find their commutator algebra:
$$\eqalign{\ \bigl[ & \ {\cal L}_{f_{1}},{\cal L}_{f_{2}}\bigr] =+i\hbar
{\cal L}_{\{f_{1},f_{2}\}_{mb}}\cr
\ \bigl[ & \ {\cal R}_{f_{1}},{\cal R}_{f_{2}}\bigr]  = -i\hbar
{\cal R}_{\{f_{1},f_{2}\}_{mb}}\cr
\ \bigl[ & \ {\cal L}_{f_{1}},{\cal R}_{f_{2}}\bigr]  =  0
\cr}\eqn\quattrododici$$
Clearly this is isomorphic to the algebra of ~$L_{f}$~and ~$R_{f}$~
of eq.~\duesette.
\par
We stressed repeatedly that the hypersurface in ~${\cal M}_{4N}$~on
which ~$\lambda=0$~is identified with the ordinary phase-space. Let us
therefore ask what is the effect of the transformations generated
by ~${\cal L}_{f}$~and ~${\cal R}_{f}$~on this surface. For any
function~$A\in Fun({\cal M}_{4N})$~we define its projection
~${\cal P}A$~by
~${\cal P}A(\lambda,\phi)=A(0,\phi)$.
Applying ~${\cal P}$~to eqs.~\quattrosette~and ~\quattrootto~we have for
(anti-)holomorphic functions:
$$\eqalign{ {\cal P}~A(Z_{-})\ast_{e} B(Z_{-}) \ = & \ A(\phi)\ast B(\phi)\cr
{\cal P}~A(Z_{+})\ast_{e} B(Z_{+}) \ = & \ B(\phi)\ast A(\phi) \cr
{\cal P}~A(Z_{-})\ast_{e} B(Z_{+}) \ = & \ {\cal P}~A(Z_{+})
\ast_{e}~B(Z_{-})=A(\phi)B(\phi)\cr}\eqn\quattroquattordici$$
These equations show  that, upon projection,
~${\cal L}_{f}~~\bigl({\cal R}_{f}\bigr)$~has a very simple effect on
holomrphic (anti-holomorphic) functions:
$$\eqalign{{\cal P}~{\cal L}_{f} g(Z_{-}) \ = & \ L_{f} g(\phi)\cr
{\cal P}~{\cal R}_{f} g(Z_{+}) \ = & \ R_{f} g(\phi)\cr}\eqn\quattroquindici$$
Interpreted in the spirit of refs.[15]~these equations tell us
how the operators ~$L_{f}$~and ~$R_{f}$~get "lifted" from
~${\cal M}_{2N}$~to ~${\cal M}_{4N}\sim T{\cal M}_{2N}$. If we go over
from the ordinary Moyal formalism to the extended one, we have to
replace ~$f(\phi)\ast$,~say, by ~$f(Z_{-})\ast_{e}$~and set ~$\lambda=0$~
after having performed the derivatives implicit in ~$\ast_{e}$.
As we mentioned already in the discussion following eq.~\quattroquattroa~,
the approach of refs.[15]~realizes the
complicated operations of the quantum exterior calculus on ~${\cal M}_{2N}$~
as the "shadow" of simpler operations on the extended phase space. In this
picture ~$L_{f}$~and ~$R_{f}$~are the "shadows"
of ~${\cal L}_{f}$~ and ~${\cal R}_{f}$.
\par
By now it should also be clear how the modular conjugation ~${\cal J}$~acts on
~${\cal M}_{4N}$. It reverses the sign of ~$\lambda_{a}$~so that
~$Z_{+}^{a}$~and ~$Z_{-}^{a}$~get interchanged\refmark{15}.
Consequently ~${\cal J}$~also interchanges ~${\cal L}_{f}$~and
~${\cal R}_{f}$,\ie, ${\cal J}{\cal L}_{f}{\cal J}={\cal R}_{f}$. 
Of course ~${\cal L}_{f}$~and ~${\cal R}_{f}$~form a representation
of ~$W_{\infty}\times W_{\infty}$~on the full space ~$Fun({\cal M}_{4N})$,
not only on the (anti-)holomorphic
subspace. However, the projection of ~${\cal L}_{f}
A(Z_{-},Z_{+})$~has no simple interpretation in terms of the ordinary
Moyal formalism: one obtains ~$L_{f}A(\phi,\phi)$~with ~$L_{f}$~
acting on the first argument of ~$A$~only.
\chapter{CONCLUSIONS}
Looking back at the previous sections in the spirit of non-commutative
geometry\Ref\conne{A.Connes, "{\it Non commutative differential
geometry}",~Publ. Math. IHES 62 (1985) 41}
, we can say that quantum phase space is a typical
example of a "non-commutative manifold". It can be studied by investigating
the properties of the algebra of functions defined over ~${\cal M}_{2N}$.
Deforming the ordinary pointwise product on ~$Fun({\cal M}_{2N})$~to the
star-product, we can introduce the Moyal bracket and obtain a quantum
deformation of the Poisson bracket. In this context "quantization" means that
we replace the infinite dimensional Lie algebra of ~$Fun({\cal M}_{2N})$~
equipped with the Poisson bracket, by the deformed algebra equipped
with the Moyal bracket. Thus the Moyal bracket algebra, like the commutator
algebra but unlike the Poisson bracket algebra, has an underlying associative
product, with respect to which it is defined as a  commutator.
All the differences between the {\it classical} and the {\it quantum}
{\it geometry} of ~${\cal M}_{2N}$~are encoded in the properties of the star-
product. By invoking the isomorphism between star-multiplication and
operator products, we have been able to relate ~$Fun({\cal M}_{2N})$~to
the GNS representation space ~${\cal V}\otimes{\cal V}$. This seems to
support the view that the quantum deformed geometry of phase-space is
indeed encoded in the GNS construction and in particular in the doubled
Hilbert space ~${\cal V}\otimes{\cal V}$. It would be nice to
study the topological and cohomological property of this space by
using the quantum exterior calculus  developed in ref.[15] and following
the lines of ref.[24]\REF\Mani{Yu.I.Manin, {\it "Notes on quantum groups
and quantum De Rham complexes"},\nextline
Max Planck Inst. f\"ur Math., Bonn, 1991}.
\par
Another issue which is  particularly intriguing is the fact that 
modular conjugation ~${\cal J}$~appears on a purely geometrical basis
here.
It owes its existence to the very nature of non-commutative geometry:
only because the star product is non-commutative there can be a difference between left
and right multiplication and hence an exchange transformation~${\cal J}$.
\ack
This reasearch has been supported in part by grants from INFN, \break
MURST and NATO.
M.R.acknowledges the hospitality of the Dipartimento Di Fisica Teorica,
Universit\`a di Trieste, while this work was in progress.
\refout
\vfil\eject
\bye